\documentclass[aps,prb,showpacs,floatfix,amsmath,amssymb,superscriptaddress,reprint]{revtex4-2}
\usepackage{float}
\usepackage{tikz}
\usepackage{bbold}
\usepackage[normalem]{ulem}
\usepackage{color}
\usepackage{graphicx}
\usepackage{epsfig}
\usepackage{setspace}
\usepackage{amsmath}
\usepackage{amssymb}
\usepackage{verbatim}
\usepackage{cancel}
\usepackage{xcolor}

\definecolor{scarred}{rgb}{0.75,0.0,0.0}
\graphicspath{ {./Images/} }
\usepackage{enumitem} 
\usepackage[colorlinks=true,citecolor=blue,linkcolor=blue,urlcolor=blue]{hyperref}

\begin{document}
\title{Universal Mott quantum criticality in a modified periodic Anderson model}
\author{Sujan K.\ K.}
\email{sujanbharadwaj97@gmail.com}
\affiliation{Jawaharlal Nehru Centre for Advanced Scientific Research, Jakkur PO, Bengaluru 560064, India.}
\author{N.\ S.\ Vidhyadhiraja}
\email{raja@jncasr.ac.in}
\affiliation{Jawaharlal Nehru Centre for Advanced Scientific Research, Jakkur PO, Bengaluru 560064, India.}

\begin{abstract}
Mott quantum criticality is a central theme in correlated electron physics, with comparable critical exponents observed in materials featuring both continuous zero-temperature transitions and those with finite-temperature critical endpoints. Such criticality was first predicted theoretically for the single-band Hubbard model (SBHM). Within dynamical mean-field theory (DMFT), the SBHM undergoes a first-order transition at $T=0$ but displays quantum critical scaling above its finite-temperature critical point. However, a comprehensive analysis of a system exhibiting a continuous Mott transition at zero temperature has been lacking. To this end, the modified periodic Anderson model (MPAM) is a rare example known to host a surface of continuous Mott quantum critical points (QCPs). While previous studies of the MPAM characterized its QCP and showed the emergence of a pseudogap Anderson model at the QCP, the analysis was restricted to the Matsubara frequency axis, leaving key questions unresolved: What are the signatures of Mott quantum criticality in transport properties, and do the critical exponents align with the universal behavior seen in the SBHM and experiments? To address these questions, we employ DMFT with the numerical renormalization group as an impurity solver to investigate the real-frequency properties of the MPAM. Our central finding is the emergence of quantum critical scaling in the electrical resistivity, with exponents $\nu z_{\text{met}} \approx 0.76$ and $\nu z_{\text{ins}} \approx 0.66$ on the metallic and insulating sides, respectively. These values fall within the range reported for the SBHM and observed in experiments, suggesting that both transitions are governed by a common universality class. We further substantiate the presence of local quantum criticality by demonstrating robust $\omega/T$ scaling in single- and two-particle correlation functions. Finally, we identify novel signatures in the optical conductivity, where the distinct evolution of two isosbestic points provides a unique fingerprint of Mott quantum criticality. These results establish the MPAM as a canonical model for investigating Mott quantum criticality and support the existence of a universal framework for this fundamental phenomenon.

\end{abstract}
\maketitle
\section{Introduction}
\label{sec:intro}
Quantum phase transitions (QPTs), which occur at absolute zero temperature when a non-thermal control parameter tunes the system between competing ground states, are of fundamental significance in the field of strongly correlated electron systems~\cite{sondhi1997continuous,sachdev1999quantum,vojta2003quantum}. A canonical example of such a phenomenon is the interaction-driven metal-to-insulator transition, widely known as the Mott transition~\cite{imada1998metal}. Extensive investigations over recent decades have revealed that this transition can be either first-order ~\cite{mcwhan1973metal,yao1996electrical,limelette2003mott,furukawa2015quantum} or continuous~\cite{neumann2007bilayer,li2021continuous,zheng2022quantum}. The single-band Hubbard model (SBHM)~\cite{hubbard1963electron,gutzwiller1963effect,kanamori1963electron}, the paradigmatic theoretical framework for this problem, correspondingly displays a first-order Mott transition terminating at a finite-temperature critical point~\cite{rozenberg1992mott,jarrell1992hubbard,georges1992numerical,rozenberg1994mott,georges1996dynamical,bulla1999zero,bulla2001finite,werner2007doping,terletska2011quantum,eisenlohr2019mott}. Intriguingly, theoretical analyses of the SBHM have shown a characteristic quantum critical scaling in the electrical resistivity above the critical temperature~\cite{terletska2011quantum,vuvcivcevic2013finite,eisenlohr2019mott}, a feature generally associated with quantum criticality. This scaling behavior was subsequently realized in experiments on materials with first-order transitions~\cite{furukawa2015quantum}, yielding a comparable scaling exponent ($\nu z$). Remarkably, a similar collapse with a consistent scaling exponent has also been observed in Moir\'{e} superlattices~\cite{li2021continuous}, which exhibit a continuous Mott transition. This convergence implies that the critical exponents derived from the SBHM may be universal, applicable even to systems that undergo a continuous Mott transition. While continuous Mott transitions have been theoretically realized in lower-dimensional or geometrically frustrated Hubbard models, such as on the 1/5-depleted square lattice~\cite{yanagi2014continuous}, a comprehensive analysis of their critical exponents is often not performed. The only other model known to exhibit a continuous ($T_c$=0) Mott transition in higher dimensions is a modified periodic Anderson model (MPAM)~\cite{sen2016quantum,kksujan2023emergent}, which describes a continuous Mott transition within the paramagnetic phase.

The MPAM is a periodic Anderson model (PAM)~\cite{georges1996dynamical,smith2003kondo,vidhyadhiraja2004dynamics,logan2016mott} coupled to an additional conduction band via an inter-orbital coupling $t_\perp$. In the particle-hole symmetric limit, the PAM is a Kondo insulator~\cite{smith2003kondo,logan2016mott}. Adding a conduction band to a PAM induces a finite spectral weight in the erstwhile localized f-band for any non-zero $t_\perp$~\cite{sen2016quantum}. Consequently, a finite interaction strength is required to drive a metal-to-insulator transition~\cite{sen2016quantum}. Indeed, the interacting MPAM~\cite{sen2016quantum,kksujan2023emergent} exhibits a quantum phase transition within the paramagnetic regime, separating a Fermi-liquid metal from a Mott insulator. At the quantum critical point (QCP), a singular density of states emerges, characterized by the formation of a soft gap~\cite{kksujan2023emergent}. This soft gap form of the hybridization is incorporated in the pseudogap Anderson model~\cite{withoff1990phase,fritz2004phase,glossop2005local,glossop2011critical,chowdhury2015critical}, which results in an impurity quantum phase transition, which is otherwise absent. In contrast, within the MPAM, this form emerges naturally within DMFT at the QCP~\cite{kksujan2023emergent}. The soft-gap spectral function of the MPAM at the QCP corresponds to the local moment fixed point of the pseudogap Anderson model. This form of the spectral function has been reported in the SBHM~\cite{eisenlohr2019mott} at the lower edge of the co-existence region ($U=U_{c1}$).

While previous studies~\cite{peters2013kondo,sen2015spectral,sen2016quantum,hu2017effects,kksujan2023emergent,majumder2025mott} have provided significant insights, the absence of finite-temperature real-frequency data leaves several fundamental questions unresolved. These include: Does the MPAM display Fermi-liquid scaling in its spectral function? How does the QCP influence the adjacent phases, as reflected in the correlation functions? Does the MPAM exhibit $\omega/T$ scaling, a hallmark of local quantum criticality? What are the key transport signatures of Mott quantum criticality? Does the resistivity exhibit a scaling collapse analogous to that observed in the SBHM, thereby suggesting a common universality class for Mott quantum criticality? To address these questions, we employ the numerical renormalization group (NRG) method within the dynamical mean-field theory (DMFT) framework~\cite{georges1996dynamical} to study the MPAM directly at finite temperatures and on the real-frequency axis.

The remainder of this paper is organized as follows: In Section~\ref{sec:Model and Methodology}, we discuss the model and methodology, where we provide a brief introduction to the MPAM and the NRG. In Section~\ref{sec:results}, we analyze the scaling of the spectra in the Fermi liquid region and examine its validity in the real frequency space. We also demonstrate the vanishing of energy scales at the QCP. Furthermore, we discuss the influence of the QCP on the self-energy and identify a crossover scale that vanishes as the QCP is approached. In Section~\ref{sec:wbyTscaling}, we present the $\omega/T$ scaling observed in the spectral function, self-energy, and dynamical spin-susceptibility at the QCP. In Section~\ref{sec:transport}, we examine the transport properties and identify two isosbestic points (ISPs) flanking the mid-infrared (MIR) peak. As the system approaches the QCP, the ISPs shift in opposite directions and eventually disappear, while the MIR peak reaches its minimum and disappears beyond the QCP. These trends serve as distinct signatures of the QCP, offering potential experimental markers for its identification. Finally, we present a key result: a characteristic quantum critical scaling in electrical resistivity, which is a hallmark of the quantum critical Mott transition. Additionally, we show that the critical scaling exponent of the MPAM closely matches that of the SBHM. This suggests the possibility of a common universality class governing Mott quantum criticality in these two models. In Section~\ref{sec:discussion}, we present our conclusions and provide an outlook on future directions.

\section{MODEL AND FORMALISM}
\label{sec:Model and Methodology}
The MPAM has a localized correlated $f$ orbital with an energy $\epsilon_f$ and on-site Coulombic repulsion term $U$. This $f$ orbital is hybridized to a conduction band $c$ with a hybridization $V$. This conduction band is coupled to another conduction band $c_M$ with an inter-orbital coupling $\text{t}_\perp$. $\epsilon_k$ is the dispersion of both conduction bands $c$ and $c_M$. The Hamiltonian of the MPAM in second quantized notation is,
\begin{align}
    H & = \sum_{\textbf{k}\sigma}\epsilon_\textbf{k} (c_{\textbf{k}\sigma}^\dagger
c_{\textbf{k}\sigma}^{\phantom{\dagger}} + c_{M\textbf{k}\sigma}^\dagger
c_{M\textbf{k}\sigma}^{\phantom{\dagger}})+ V \sum_{\textbf{k}\sigma}( f^\dag_{\textbf{k}\sigma}c_{\textbf{k}\sigma}^{\phantom{\dagger}} + {\rm h.c})  \nonumber \\
 & \qquad+ t_\perp \sum_{\textbf{k}\sigma}( c^\dag_{\textbf{k}\sigma}c_{M\textbf{k}\sigma}^{\phantom{\dagger}} + {\rm h.c}  )+\epsilon_f\sum_{i\sigma} f^\dag_{i\sigma}f^{\phantom{\dag}}_{i\sigma} \nonumber \\
 &\qquad + U\sum_i n_{fi\uparrow}n_{fi\downarrow}\label{eq:mpam}  \,.
\end{align}

The MPAM in the particle–hole symmetric limit has been examined within DMFT in the paramagnetic regime, using the continuous-time quantum Monte Carlo (CT-QMC) method at finite temperatures ~\cite{kksujan2023emergent} and the local moment approach (LMA) at zero temperature ~\cite{sen2016quantum} as impurity solvers. Both studies demonstrate that the model exhibits a surface of QCPs in the $U-t_\perp$ plane, separating the Fermi liquid (FL) phase from the Mott insulating phase. Throughout this work, we fix the hybridization at $V=0.44$ and restrict ourselves to the paramagnetic, particle-hole symmetric phase, which corresponds to setting $\epsilon_c = \epsilon_M = 0$ and $\epsilon_f = -U/2$. The chemical potential is set to zero throughout.

In this work, we solve the lattice model within the framework of dynamical mean-field theory (DMFT). DMFT is a powerful non-perturbative approach that maps a lattice problem onto an effective single-impurity model, which is then solved self-consistently ~\cite{metzner1989correlated,georges1996dynamical,vollhardt2011dynamical}. This mapping becomes exact in the limit of infinite spatial dimensions and neglects all nonlocal contributions to the retarded self-energy. Consequently, the retarded self-energy is purely local, i.e., $\Sigma_{\text{f}}(\omega,\mathbf{k}) \equiv \Sigma_{\text{f}}(\omega)$. The self-consistency condition for the MPAM within the DMFT framework is expressed as
\begin{align}
    G_{\text{loc}}(\omega)&=\int\,d\epsilon\,\text{A}_{\text{0}}(\epsilon)\,G_{f}(\omega,\epsilon) \nonumber\\&= \frac{1}{\omega^+-\epsilon_f-\Delta(\omega)-\Sigma_{\text{f}}(\omega)}\label{eq:self_consistancy}
\end{align}
Here, the retarded $f-$Green's function is given as,
\begin{align*}
        G_{f}^{-1}(\omega,\epsilon) &= \omega^+ + \mu - \Sigma_{\text{f}}(\omega) - \frac{V^2}{\omega^+ - \epsilon - \frac{t^{2}_{\perp}}{\omega^+-\epsilon }},
\end{align*}
where $\text{A}_{\text{0}}(\epsilon)=2\sqrt{1-\epsilon^2/D^2}/\pi D$ is a semi-elliptic non-interacting conduction band density of states, with a half-bandwidth $D=1$ used throughout this work. $\Sigma_{\text{f}}(\omega)$ denotes the momentum-independent local retarded self-energy, and $\Delta(\omega)$ represents the hybridization function of the effective bath. Similarly, for later use, the retarded Green's functions corresponding to the two metallic orbitals and two mixed orbitals, denoted as $G_{\text{cc}}(\omega,\epsilon)$, $G_{\text{mm}}(\omega,\epsilon)$,$G_{\text{cm}}(\omega,\epsilon)$ and $G_{\text{mc}}(\omega,\epsilon)$  are given by:
\begin{align}
    G_{\text{cc}}(\omega,\epsilon)&= \left[\omega^+ - \epsilon - \frac{V^2}{\omega^+ + \mu - \Sigma_{\text{f}}(\omega)} - \frac{t_{\perp}^2}{\omega^+ - \epsilon}\right]^{-1}\label{eq:Gcc} \\
    G_{\text{mm}}(\omega,\epsilon)& = \left[\omega^+ - \epsilon - \frac{t_{\perp}^2}{\omega^+ - \epsilon - \frac{V^2}{\omega^+ + \mu - \Sigma_{\text{f}}(\omega)}}\right]^{-1}\label{eq:Gmm}\\
    G_{\text{mc}}(\omega,\epsilon) &= G_{\text{cm}}(\omega,\epsilon) = \frac{t_\perp}{\omega^+-\epsilon}\,G_{\text{cc}}(\omega,\epsilon)\label{eq:Gcm}
\end{align}
Here, the Green's functions are defined in standard Zubarev~\cite{zubarev1960double,vidhyadhiraja2007specific} notation as:
\begin{align*}
    G_{\text{cc}}(\omega,\epsilon)\equiv\langle\langle c_{\textbf{k}\sigma};c_{\textbf{k}\sigma}^\dagger\rangle\rangle_\omega,\, G_{\text{mm}}(\omega,\epsilon)\equiv\langle\langle c_{M\textbf{k}\sigma};c_{M\textbf{k}\sigma}^\dagger\rangle\rangle_\omega,\\
    G_{\text{mc}}(\omega,\epsilon)\equiv \langle\langle c_{M\textbf{k}\sigma};c_{\textbf{k}\sigma}^\dagger\rangle\rangle_\omega, G_{\text{cm}}(\omega,\epsilon)\equiv \langle\langle c_{\textbf{k}\sigma};c_{M\textbf{k}\sigma}^\dagger\rangle\rangle_\omega.
\end{align*}

To satisfy the equality in Eq.~\ref{eq:self_consistancy}, the impurity problem must be solved iteratively using the self-consistently determined hybridization function at each step. The impurity problem can be solved using a variety of methods, including Iterative Perturbation Theory (IPT)~\cite{yamada1975,yosida1975perturbation,kajueter1996new}, the numerically exact Continuous-Time Quantum Monte Carlo (CTQMC) method~\cite{rubtsov2004continuous,rubtsov2005continuous,werner2006hybridization,gull2008continuous}, the Local Moment Approach (LMA)~\cite{logan1998local,vidhyadhiraja2004dynamics,galpin2005single}, and the Numerical Renormalization Group (NRG)~\cite{wilson1975renormalization}. Each of the methods has its own drawbacks. For example, IPT is a second order approximation theory and fails to capture non-perturbative effects. CTQMC is numerically exact but gives results on the Matsubara axis. To obtain real frequency data, it is necessary to perform analytic continuation, an ill-defined procedure. Furthermore, CTQMC calculations get prohibitively expensive, particularly at very low temperatures. On the other hand, NRG is a non-perturbative technique that works in real frequencies. To address the impurity problem within DMFT, we employ the NRG method, utilizing the open-source implementation developed by Rok Žitko ~\cite{vzitko2009energy}.

The NRG, a non-perturbative technique, was originally developed by Wilson ~\cite{wilson1975renormalization}. To accurately capture all energy scales, the conduction bath in NRG is discretized into logarithmic intervals $[{\Lambda^{n+1}, \Lambda^n}]$ ($n = 0, 1, 2, \ldots$), where $\Lambda > 1$ is the discretization parameter. The width of each interval is given by $d_n = \Lambda^{-n}(1 - \Lambda^{-1})$, and the continuum limit is recovered as $\Lambda \rightarrow 1$. The continuum of conduction band states is then replaced by a discrete set of states, which is mapped onto a semi-infinite tight-binding chain with the impurity located at one end. The hopping amplitudes along the chain decay as $\Lambda^{-n/2}$. The resulting Hamiltonian is solved iteratively, but the exponential growth of the Hilbert space with system size presents a computational challenge. To address this, Wilson introduced a truncation scheme~\cite{wilson1975renormalization} in which the number of retained states is limited to $N_{\text{states}}$, as the discarded high-energy states have negligible influence on the low-energy physics. The NRG has proven highly successful in the study of quantum impurity problems ~\cite{krishna1980renormalization,sakai1989single,costi1994transport,bulla1998numerical,bulla2008numerical}, and has also been widely employed as an impurity solver within DMFT for lattice models such as the Hubbard model ~\cite{sakai1994application, bulla1999zero}, the PAM ~\cite{pruschke2000low}, and others. In this work, we adopt a discretization parameter of $\Lambda = 2$ and perform twist averaging over $N_z = 16$ ~\cite{campo2005alternative}, whereby separate calculations are carried out for different interleaved discretization grids, and the resulting quantities are subsequently averaged. To accelerate the convergence of DMFT iterations, we employ Broyden’s mixing scheme ~\cite{vzitko2009convergence}.

\section{RESULTS}
\label{sec:results}
\subsection{Analysis of single particle quantities}
\label{subsecA}
In this section, we demonstrate universal FL scaling, the continuous vanishing of the quasiparticle weight and Mott gap at the QCP, and identify the crossover scale to the quantum critical region from the self-energy.
Understanding universal behavior across different phases is crucial, as it provides deeper insight into the nature of the phases and facilitates comparison with experimental results~\cite{vidhyadhiraja2005optical}. In the FL regime of the MPAM, and in the limit $\omega \rightarrow 0$, the single-particle $f$-spectral function scales as $\text{A}_{\text{f}}(\omega) =-\text{Im}[G_{\text{loc}}(\omega)]/\pi= (Z^2V^2/\omega^2)\text{A}_{\text{c}}(\omega)$~\cite{sen2016quantum}, where $Z= \left(1 - \frac{d}{d\omega} \text{Re}\, \Sigma_{\text{f}}(\omega) \right)^{-1}$ denotes the quasiparticle weight and $\text{A}_{\text{c}}(\omega)=\text{-Im}[G_{\text{c}}(\omega)]$ $\left(G_\text{c}(\omega)=\int\,d\epsilon\,G_\text{c}(\omega,\epsilon)\right)$ is the $c$-band spectral function.
In the low-frequency limit ($\omega\rightarrow0$), the c-band spectral function (see Appendix~\ref{sec:Appendix2} for the derivation) is given by: 
\begin{align*}
    &\text{A}_{\text{c}}(\omega)\overset{\omega \rightarrow 0}{\sim} \left(\frac{\omega t_\perp}{ZV^2}\right)^2\text{A}_{\text{0}}\left[\omega\left(1+\frac{t_{\perp}^2}{ZV^2}\right)\right]\\
    &\qquad+\left[1-\left(\frac{\omega t_\perp}{ZV^2}\right)^2\right]\text{A}_{\text{0}}\left[\omega\left(1-\frac{t_{\perp}^2}{ZV^2}\right)-\frac{ZV^2}{\omega}\right].
\end{align*}
As $\omega\rightarrow0$ and for $ZV^2\ll t_{\perp}^2$, the first term dominates, and in this regime, the scaled $f$-electron spectral function is given by $\Tilde{\text{A}}_\text{f}(\omega) \equiv \text{A}_{\text{f}}(\omega)(V^2/t_{\perp}^2) = \text{A}_{\text{0}}(\Tilde{\omega})$, where the rescaled frequency is defined as $\Tilde{\omega} \equiv \omega/\omega_L$, with $\omega_L= ZV^2/t_{\perp}^2$. The energy scale $\omega_L$  is the counter part of the Kondo scale~\cite{hewson1997kondo} in the Anderson impurity model and coherence scale in the periodic Anderson model~\cite{vidhyadhiraja2004dynamics}. In this limit, the spectral function is expected to have universal scaling of the form $\mathcal{F}(\omega/\omega_L,T/\omega_L)(\omega_L=ZV^2/t_{\perp}^2)$~\cite{sen2016quantum,vidhyadhiraja2004dynamics}.

\begin{figure}[htbp]
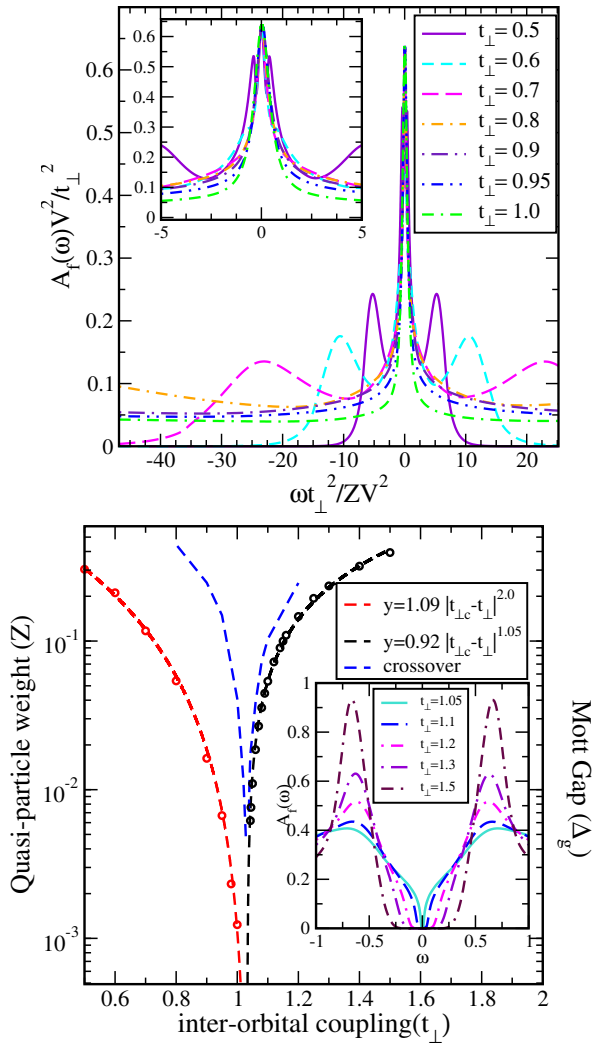

    \centering
    \includegraphics[width=0.8\linewidth, trim=2 0 0 0, clip]{Figure1a.eps}\\[0.5ex]
    \includegraphics[width=0.9\linewidth, trim=0 0 0 0, clip]{Figure1b.eps}
    \caption{(Top) Rescaled spectral weight, $\text{A}_{\text{f}}(\omega)V^2/t_{\perp}^2$, as a function of $\omega t_{\perp}^2/ZV^2$ at $T=10^{-3}$. 
    (Bottom) Energy scales: quasi-particle weight (${Z}$)(solid lines- red), crossover scale(dashed-blue) and Mott gap ($\Delta_g$)(solid lines- black) vs. inter-orbital coupling $t_\perp$. The crossover scale is extracted from the self-energy deviation from the power-law form(see text corresponding to Fig. ~\ref{fig:self_energy_wm}). $Z$ and $\Delta_g$, calculated at $T=0$, vanish with exponents $2.0$ and $1.05$. Inset: $\text{A}_{\text{f}}(\omega)$ for different $t_\perp$. Parameters used: $U=1.75$.}
    \label{fig:spectra}
\end{figure}

\begin{figure}[htbp]
    \centering
    \includegraphics[width=0.48\textwidth,trim={0 0 0 1},clip]{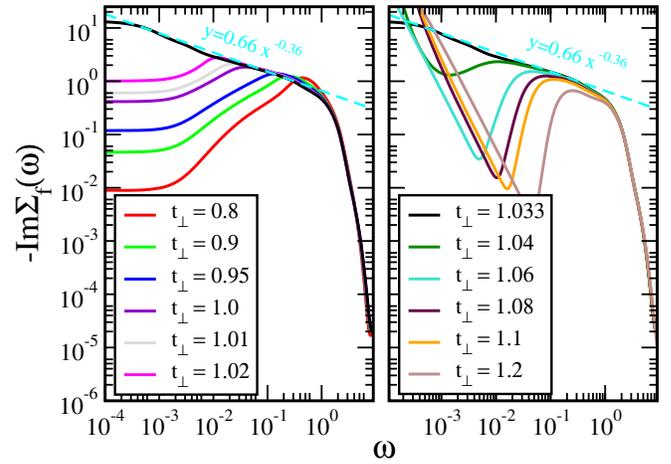}
    \caption{-Im$\Sigma_{\text{f}}(\omega)$ is plotted for different inter-orbital coupling, $t_\perp$ at a temperature $T=0.001$. We can observe a gradual influence of the power-law as we approach the QCP. Parameters used: $U=1.75$.}
    \label{fig:self_energy_wm}
\end{figure}
The top panel of Fig.~\ref{fig:spectra} displays the scaled spectral function in the FL regime. As expected, the spectral function exhibits a scaling collapse as $\omega \rightarrow 0$ and remains pinned at $\omega = 0$. This observation confirms the principle of adiabatic continuity, a necessary condition for the existence of a Fermi liquid phase. In the Mott insulating phase, the Hubbard bands move further apart with increasing $t_\perp$ (see the inset of the bottom panel in Fig.~\ref{fig:spectra}). To quantify the energy scales, we plot both the quasiparticle weight $Z$ and the Mott gap $\Delta_g$ calculated in the limit $T\rightarrow0$ as functions of the inter-orbital coupling $t_\perp$ (see the bottom panel of Fig.~\ref{fig:spectra}). For $t_\perp < t_{\perp,c}$, the quasiparticle weight vanishes continuously as $t_\perp \rightarrow t_{\perp,c}^{-}$ with an exponent of approximately $2.0$, which is close to the DMFT+LMA~\cite{sen2016quantum} result of $\approx 2.5$. In the insulating phase, the Mott gap $\Delta_g$ vanishes at the transition with an exponent of $\approx 1.05$, which is in good agreement with the value $\approx 1.0$ obtained using DMFT+LMA~\cite{sen2016quantum}. 

Our calculated exponents, while consistent with DMFT+LMA~\cite{sen2016quantum} results, differ from those obtained using DMFT+CTQMC~\cite{kksujan2023emergent}. This discrepancy can be attributed to the distinct methodologies for determining the quasiparticle weight, $Z$. The previous CTQMC study approximated $Z$ from the first Matsubara frequency point, with this value then extrapolated to $T=0$. In contrast, our approach leverages the NRG solver to compute the real-frequency self-energy. This enables a direct calculation of $Z$ from its derivative at zero frequency. Beyond low-energy scales, we investigate whether an additional scale characterizes the quantum critical region. To explore this, we analyze the self-energy.

The imaginary part of the self-energy in the FL phase varies as $-(\omega^2 + \pi^2 T^2)$. In contrast, in the Mott insulating (MI) phase, $\text{Im}\,\Sigma_{\text{f}}(\omega)$ diverges. Due to scale invariance at the QCP, $\text{Im}\,\Sigma_{\text{f}}(\omega)$ also diverges at criticality. Fig.~\ref{fig:self_energy_wm} shows the plot of $-\text{Im}\,\Sigma_{\text{f}}(\omega)$ for various values of $t_\perp$. In the FL regime ($t_\perp \ll t_{\perp,c}$), $-\text{Im}\,\Sigma_{\text{f}}(\omega)$ exhibits a quadratic behavior in $\omega$, while in the MI regime ($t_\perp \gg t_{\perp,c}$), it diverges. At the critical point $t_{\perp,c} = 1.033$, $-\text{Im}\,\Sigma_{\text{f}}(\omega)$ follows a power-law behavior with a diverging exponent of $-0.35$, as found in Ref.~\citenum{kksujan2023emergent}. 
As we move away from the QCP, remnants of the power-law behavior remain visible in both the FL and MI regions. We identify the frequency at which $-\text{Im}\,\Sigma_{\text{f}}(\omega)$ begins to deviate from the power-law form on both sides as a crossover scale. The region between these crossover scales represents the quantum critical region. This scale is plotted as a function of $t_\perp$ in the bottom panel of Fig.~\ref{fig:spectra}. The crossover frequency scale vanishes as the system approaches the QCP. The proximity of the crossover energy scale to the Mott gap on the insulating side of the transition indicates the absence of any intermediate phases between the quantum critical region and the Mott insulator. However, in the FL region, the crossover scale is well separated from the Fermi liquid scale, ${Z}$. This indicates that there is a bad metal region between the quantum critical region and the FL region, similar to the one found in the SBHM~\cite{dasari2017quantum}. In the next section, we examine the quantum critical region and its scaling behavior.

\subsection{\texorpdfstring{$\omega/T$}{omega/T} scaling at the QCP:}
\label{sec:wbyTscaling}
\begin{figure}[htbp]
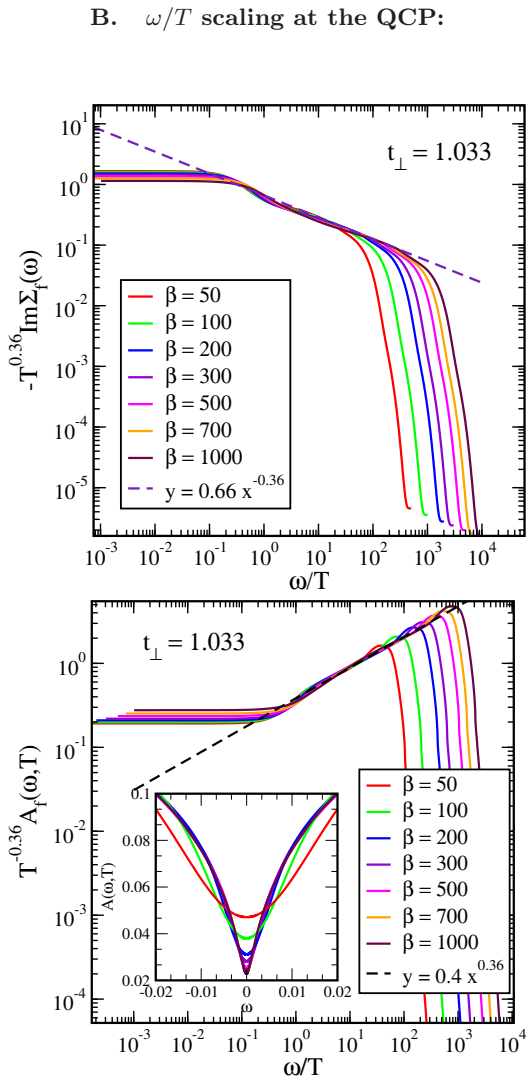

    \centering
    \includegraphics[width=0.8\linewidth,trim={0 0 0 1},clip]{Figure3a.eps}\\[0.5ex]
    \includegraphics[width=0.8\linewidth,trim={0 0 0 1},clip]{Figure3b.eps}
\caption{(top)Imaginary part of the self-energy, $-\Sigma_{\text{f}}(\omega)$ and (bottom)spectral function, $\text{A}_{\text{f}}(\omega)$, are plotted for different $\beta$'s at the QCP. We can observe $\omega/T$ scaling in both of them with  exponents $=0.36 \,(\text{A}_{\text{f}}(\omega)),-0.36\,(\text{Im}\Sigma_{\text{f}}(\omega))$. (Inset) shows the plot of unscaled $A(\omega)$ vs. $\omega$. Parameters used: $U=1.75$.}
\label{fig:qcp_nrg}
\end{figure}

\begin{figure}[htbp]
    \centering
    \includegraphics[width=0.9\linewidth,trim={0 0 0 1},clip]{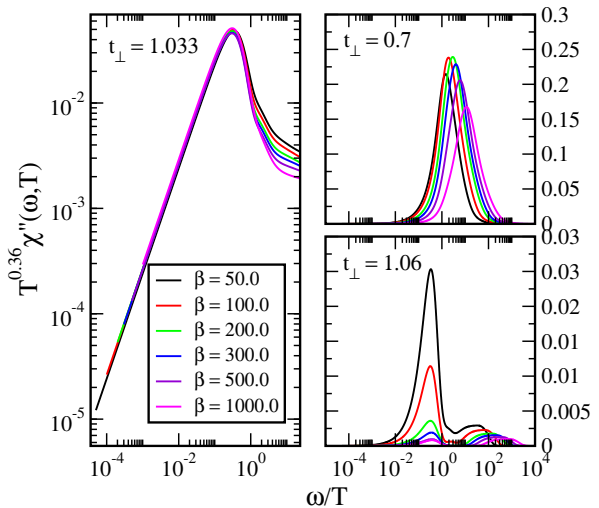}
    \caption{Scaling plot of the dynamical susceptibility, $T^{0.36}\chi''(\omega,T)$, shown for several temperatures ($\beta$ ranging from 50 to 1000). (a) At the quantum critical point ($t_{\perp} = t_{\perp,c} = 1.033$), the data collapse onto a universal curve over three decades, consistent with quantum critical scaling~\cite{glossop2011critical}. (b, c) Away from the QCP, for $t_\perp = 0.7$ (top right) and $t_\perp = 1.1$ (bottom right), the scaling collapse clearly fails. Parameters used: $U=1.75$.}
    \label{fig:wbyT_susceptibility}
\end{figure}

The SBHM~\cite{dasari2017quantum}, Kondo lattice model~\cite{si2003local}, pseudogap Anderson model~\cite{glossop2011critical}, and the pseudogap Kondo model display ``local" quantum criticality and are often associated with the $\omega/T$ scaling in the quantum critical region. Numerous neutron scattering experiments have reported quantum criticality manifested through $\omega/T$ scaling ~\cite{aronson1995non,stockert1998two,schroder2000onset,michon2023reconciling}, which belong to the class of ``local" quantum criticality~\cite{si2001locally,ingersent2002critical,si2003local,gegenwart2008quantum}. In this section, we investigate the presence of such $\omega/T$ scaling behavior in the MPAM.

As a first step towards identifying such signatures, we analyze the single-particle dynamics~\cite{glossop2011critical}. Fig.~\ref{fig:qcp_nrg} shows $\omega/T$ scaling in both the scaled spectral function, $T^{-r}\text{A}_{\text{f}}(\omega)$, and the scaled imaginary part of the self-energy, $-T^r\,\text{Im}\,\Sigma_{\text{f}}(\omega)$. The scaling exponent is found to be $r = 0.36$, consistent with the value reported in Ref.~\citenum{kksujan2023emergent}. The inset in the bottom part of Fig.~\ref{fig:qcp_nrg} displays the spectral function for various temperatures at the QCP.

A more definitive hallmark of local quantum criticality is the observation of $\omega/T$ scaling in the dynamical spin-spin susceptibility, $\chi(\omega,T)$.  
The dynamical spin susceptibility within DMFT describes the local magnetic response of the impurity to a time-dependent magnetic field at frequency $\omega$. In real frequency, we can write:
\begin{align}
\chi_{\text{loc}}^{R}(\omega) = -i \int_{0}^{\infty} dt \, e^{i \omega t} \, \langle [ S^{z}(t), S^{z}(0) ] \rangle_{\text{imp}},
\end{align}
where $S^{z} = \frac{1}{2}(n_{\uparrow} - n_{\downarrow})$ is the local spin operator.
To this end, we examine the two-particle response by computing the scaled imaginary part of the dynamical spin-spin susceptibility, $T^{r}\chi''(\omega,T)(r=0.36)$, where $\chi''(\omega,T)$ denotes the imaginary part of $\chi(\omega,T)$. At the QCP ($t_\perp = 1.033$), we observe clear $\omega/T$ scaling in $T^{0.36}\chi''(\omega,T)$, as shown in Fig.~\ref{fig:wbyT_susceptibility}. Importantly, such scaling is observed only at the critical point; as the system is tuned away from the QCP, the scaling behavior disappears as shown in the Fig. ~\ref{fig:wbyT_susceptibility}. 

The demonstration of robust $\omega/T$ scaling in both single- and two-particle correlation functions establishes that the QCP in the MPAM is locally critical. Building on this finding, we now turn to the transport properties of the MPAM to identify the signatures of Mott quantum criticality.

\subsection{Transport quantities}
\label{sec:transport}
In this section, we analyze the optical conductivity to identify signatures of the Mott quantum criticality. We further compute the DC conductivity and resistivity, demonstrating a scaling collapse of the latter with an exponent comparable to that of the SBHM.

\subsubsection{Optical conductivity}

\begin{figure}[htbp]
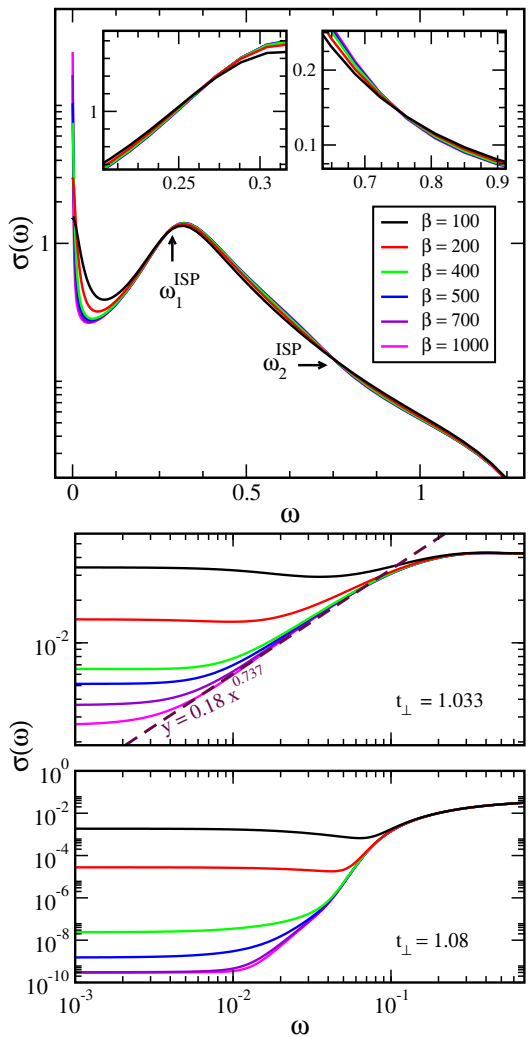

    \centering
    \includegraphics[width=0.8\linewidth,trim={0 0 0 1},clip]{Figure5a.eps}\\[0.5ex]
    \includegraphics[width=0.8\linewidth,trim={0 0 0 1},clip]{Figure5b.eps}
    \caption{(top) Optical conductivity plotted for various inverse temperatures $\beta$ at $t_\perp = 0.7$. The inset shows two isosbestic points located before and after the MIR peak. (bottom) Optical conductivity plotted for $t_\perp = 1.033$ (top) and $1.08$ (bottom). At the QCP, the conductivity $\sigma(\omega)$ follows a power-law behavior, as indicated by the dashed maroon line, with an exponent of $0.737$, which is approximately $2r$, where $r = 0.36$ is the critical exponent~\cite{kksujan2023emergent}. Parameters used: $U=1.75$.}
    \label{fig:opt_cond_nrg}
\end{figure}
A general expression for the optical conductivity, based on the Kubo formula, is given by $\displaystyle\sigma(\omega) = \frac{1}{i\omega^+} \, \langle\langle j ; j \rangle\rangle_\omega,$
where $\langle\langle j ; j \rangle\rangle_\omega$ denotes the retarded current-current correlation function. Within DMFT, the self-energy is purely local, and vertex corrections vanish in the skeleton expansion of the current–current correlation function. Consequently, only the bare bubble diagram contributes to the conductivity, which implies that single-particle quantities are sufficient to compute \(\sigma(\omega)\)~\cite{georges1996dynamical}. The optical conductivity is then calculated using the standard expression~\cite{georges1996dynamical,terletska2011quantum,vuvcivcevic2013finite,eisenlohr2019mott}:
\begin{align}
\frac{\sigma(\omega)}{\sigma_0} = 2\pi \int d\epsilon \,\text{A}_0(\epsilon) \int d\omega' \frac{f(\omega') - f(\omega'+\omega)}{\omega}\nonumber\\ \times\, \text{Tr}[v^2(\epsilon) \textbf{A}(\omega',\epsilon) \textbf{A}(\omega'+\omega,\epsilon)]\label{eq:opt_cond_w}
\end{align}
where $\mathbf{A}(\omega,\epsilon)$ is a $2\times2$ spectral function matrix with diagonal elements -Im$G_{\text{cc}}(\omega,\epsilon)$, -Im$G_{\text{mm}}(\omega,\epsilon)$, and off-diagonal elements -Im$G_{\text{mc}}(\omega,\epsilon)$, -Im$G_{\text{cm}}(\omega,\epsilon)$. $\text{A}_0$ is the non-interacting Bethe lattice density of states~\cite{vuvcivcevic2013finite,vzitko2015repulsive,eisenlohr2019mott}, $v(\epsilon) = \sqrt{D^2 - \epsilon^2}/\sqrt{3}$ represents the energy-dependent velocity on the Bethe lattice~\cite{vzitko2015repulsive,eisenlohr2019mott}, $n_F(\omega) = (1 + e^{\beta \omega})^{-1}$ is the Fermi–Dirac distribution, and $\sigma_0$ is a material-dependent constant. This expression holds in the $z \rightarrow \infty$ limit of the Bethe lattice.

\begin{figure}[htbp]
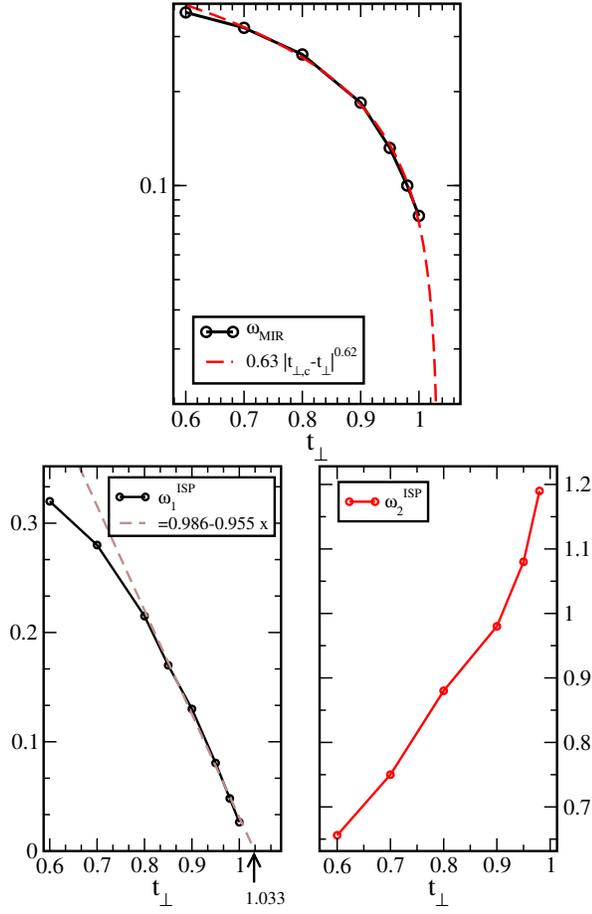

    \centering
    \includegraphics[width=0.5\linewidth,trim={0 0 0 1},clip]{Figure6a.eps}
    \includegraphics[width=0.9\linewidth,trim={0 0 0 1},clip]{Figure6b.eps}\\
    \caption{Top: Position of the MIR peak ($\omega_{\text{MIR}}$) in the optical conductivity \textit{vs.} $t_\perp$, showing that the MIR peak vanishes at the QCP. Bottom: Frequencies of the first ($\omega^{\text{ISP}}_{1}$, left panel) and second ($\omega^{\text{ISP}}_{2}$, right panel) isosbestic points in the optical conductivity as functions of $t\perp$. The first isosbestic point ($\omega^{\text{ISP}}_{1}$) vanishes at the QCP, whereas the second ($\omega^{\text{ISP}}_{2}$) disappears beforehand.Parameters: $U = 1.75$.}
    \label{fig:isosb}
\end{figure}
The optical conductivity is computed for various values of inter-orbital coupling $t_\perp$ and inverse temperature $\beta$, as shown in Fig.~\ref{fig:opt_cond_nrg}. For $t_\perp = 0.7$, we observe three distinct features: (1) a low-frequency Drude peak, originating from coherent quasiparticle excitations near the Fermi level; (2) a mid-infrared peak arising from transitions between the lower Hubbard band and the quasiparticle resonance, and (3) a high-energy charge-excitation peak corresponding to transitions between the lower and upper Hubbard bands. With decreasing temperature, the Drude peak becomes narrower and its height increases, whereas the charge-excitation peak remains largely temperature independent, similar to the earlier reports in the doped infinite-dimensional Hubbard model~\cite{jarrell1995optical}. As $t_\perp$ approaches the critical value $t_{\perp,c}$, the mid-IR peak is suppressed. For $t_\perp > t_{\perp,c}$, an optical gap develops due to the absence of coherent quasiparticles.

The three-peak structure in $\sigma(\omega)$ for $t_\perp \ll t_{\perp,c}$ arises due to different interband excitations. This can be understood by analyzing the roots of the Green’s function $G(\omega,\epsilon)$ in the following way. Assuming that the Fermi liquid ground state is adiabatically connected to the non-interacting limit with a quasiparticle renormalization factor $Z$, the spectral function becomes a sum of delta functions: $\text{A}(\omega,\epsilon) \propto \sum_{i=u,m,l} \delta(\omega - R_i(\epsilon))$, where $R_u$, $R_m$, and $R_l$ are roots corresponding to the upper, middle, and lower bands, respectively. In the limit $|\omega| \gg \frac{ZV^2}{2t_\perp}$, the roots of the Green’s function can be approximately given as (see Appendix~\ref{sec:Appendix1} for details):
\begin{align*}
    R_u &= \frac{\epsilon + t_\perp + \sqrt{(\epsilon + t_\perp)^2 + 2ZV^2}}{2},\\
    R_m &= \frac{\epsilon - \text{sgn}(\epsilon)\, t_\perp + \text{sgn}(\epsilon)\sqrt{(\epsilon - \text{sgn}(\epsilon)t_\perp)^2 + 2ZV^2}}{2},\\
    R_l &= \frac{\epsilon - t_\perp - \sqrt{(\epsilon - t_\perp)^2 + 2ZV^2}}{2}.
\end{align*}

In the metallic regime, the optical conductivity spectrum for $\omega \geq 0$ is composed of three distinct features. The first is a Drude peak at zero frequency. The other two features arise from three possible intraband transitions: $\Delta_{ul}$, $\Delta_{um}$, and $\Delta_{ml}$. Of these, the $\Delta_{um}$ and $\Delta_{ml}$ transitions are degenerate, combining to form a single mid-infrared (MIR) peak. The remaining transition, $\Delta_{ul}$, gives rise to the charge excitation peak. Consequently, the total optical spectrum clearly exhibits these three peaks. The charge-excitation peak in $\sigma(\omega)$, corresponding to the transition from the lower to the upper band, is minimum when $\epsilon=0$, and is approximately given by $\Delta_{ul}(\epsilon=0) = t_\perp + \sqrt{t_\perp^2 + 2ZV^2} \approx 2t_\perp$. It is interesting to note that, in the limit $t_\perp = 0$, the MPAM reduces to the PAM, and $\Delta_{ul} \approx \sqrt{2ZV^2}$, which is smaller by a factor of $\sqrt{2}$ compared to the gap obtained for the PAM~\cite{logan2005dynamics} ($= 2\sqrt{ZV^2}$). This discrepancy arises from the approximation used in determining the roots of the MPAM (see Eq.~\ref{eq:A1}), where the term $\frac{ZV^2}{2\omega t_\perp}$ is neglected. Including this term would yield the same result as the PAM in the limit $t_\perp \rightarrow 0$.

\begin{figure}[htbp]
    \centering
    \includegraphics[width=0.8\linewidth,trim={0 0 0 1},clip]{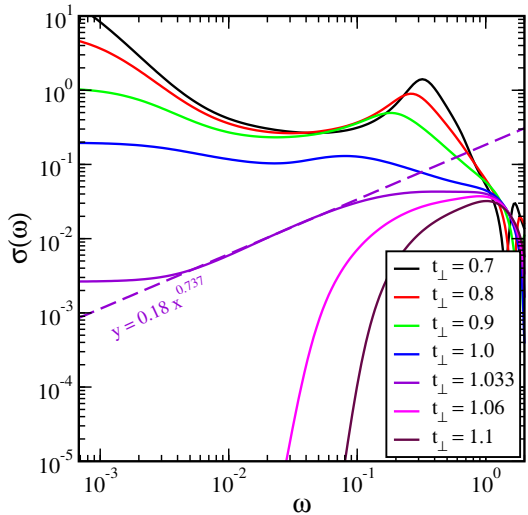}
    \caption{Optical conductivity plotted for $\beta = 10^3$ and different $t_\perp$ values. The MIR peak shifts toward $\omega = 0$ as the QCP ($t_\perp = t_{\perp,c}$) is approached and disappears at this point. Parameters used: $U=1.75$.}
    \label{fig:optcond_difftp}
\end{figure}

The MIR peak originates from either $\Delta_{um}$ or $\Delta_{ml}$. Their contributions are identical due to particle–hole symmetry, with the minimum excitation occurring at $\epsilon = -\min(D, t_\perp)$ for $\Delta_{um}$ and at $\epsilon = \min(D, t_\perp)$ for $\Delta_{ml}$. Hence, the position of the MIR peak, $\omega_{\text{MIR}}$, is proportional to $\Delta_{um}(\epsilon = -\text{min}(D, t_\perp)) = \Delta_{ml}(\epsilon = \text{min}(D, t_\perp)) \approx \sqrt{|\text{min}(D, t_\perp) - t_\perp|^2 + 2ZV^2}$. As the system approaches the critical point, both $Z \rightarrow 0$ and $|\text{min}(D, t_\perp) - t_\perp|$ decrease. Consequently, the MIR peak redshifts towards $\omega = 0$ and the charge-excitation peak blue-shifts away from $\omega = 0$ as the system approaches the QCP. In particular, $\omega_{\text{MIR}}$ reaches its minimum at the QCP with an exponent of $0.62$, as shown in the top panel of Fig.~\ref{fig:isosb}. Notably, the MIR peak persist even in the $Z \rightarrow 0$ limit due to the finite contribution from $|D - t_{\perp,c}|$ ($t_{\perp,c} = 1.033 > D$). This trend is evident in Fig.~\ref{fig:optcond_difftp}, where the MIR peak moves towards $\omega = 0$ as $t_\perp$ increases, reaching a minimum at $t_\perp = t_{\perp,c}$. For $t_\perp < D = 1.0$, however, $\omega_{\text{MIR}}\propto \sqrt{2ZV^2}$, which is explicitly independent of $t_\perp$. This implies that the MIR peak provides information about the hybridization gap and quasi-particle weight, similar to the case in the PAM~\cite{vidhyadhiraja2005optical}. However, as $t_\perp \rightarrow t_{\perp,c}$, other model parameters (here, $t_\perp$) can also influence the MIR peak. Beyond the QCP ($t_\perp>t_{\perp c}$), the MIR peak disappears. At the QCP, due to the diverging nature of $\Sigma(\omega)$, the spectral functions, $A(\omega,\epsilon)\propto\omega^r$, with $r=0.36$~\cite{kksujan2023emergent} being the soft-gap exponent. This results in a power-law behavior of the optical conductivity with an exponent $2r$.
The bottom panel of Fig.~\ref{fig:opt_cond_nrg} shows that as the temperature is decreased, the optical conductivity approaches a power-law form with an exponent $0.737$, which is approximately $2r$. 

Upon examining the top panel of Fig.~\ref{fig:opt_cond_nrg}, we identify two crossing points, commonly referred to as isosbestic points (ISPs)~\cite{greger2013isosbestic}. In the present context, these correspond to temperature-independent crossings in the optical conductivity for a given set of parameters. Similar isosbestic behavior has been reported previously~\cite{greger2013isosbestic,jarrell1995optical}. In the regime $t_\perp < t_{\perp,c}$, two ISPs are clearly visible: one located below the mid-infrared (MIR) peak ($\omega_1^{\text{ISP}} < \omega_{\text{MIR}}$) and another above it ($\omega_2^{\text{ISP}} > \omega_{\text{MIR}}$). These ISPs gradually disappear as the system approaches the QCP. We track them by extracting the corresponding frequencies, $\omega_{1/2}^{\text{ISP}}$. The bottom panel of Fig.~\ref{fig:isosb} presents the evolution of ISPs as a function of $t_\perp$. We observe that $\omega_1^{\text{ISP}}$ vanishes linearly at the QCP, while $\omega_2^{\text{ISP}}$ disappears before reaching the QCP. Observing these distinct behaviors experimentally would offer compelling evidence for the existence of a QCP.

Our analysis of transport properties reveals clear signatures of Mott quantum criticality in the optical conductivity. We now turn to the DC resistivity to investigate its quantum critical scaling behavior.

\subsubsection{DC conductivity and resistivity scaling}

\begin{figure}[htbp]
    \centering
    \includegraphics[width=0.8\linewidth,trim={0 0 0 1},clip]{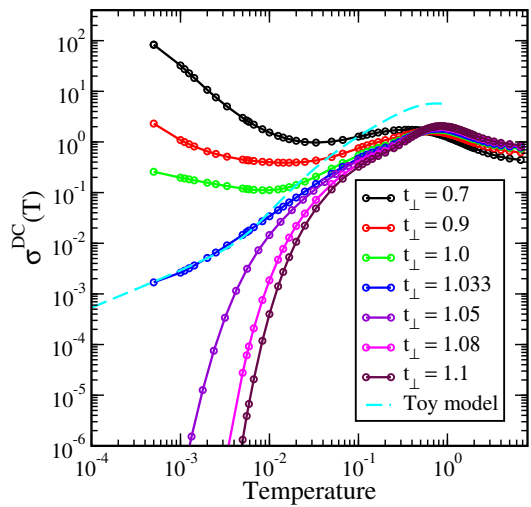}\\
    \caption{The DC conductivity, $\sigma^{\text{DC}}(T)$, computed using Eq.~\ref{eq:dc_cond}, is plotted for several values of the inter-orbital coupling $t_\perp$. The blue dashed line denotes the result from a toy model calculation at the QCP, obtained by substituting the power-law form of the self-energy, $\Sigma_{\text{f}}(\omega) \propto T^{-r}|\omega/T|^{-r}$, into Eq.~\ref{eq:dc_cond}. As $T \rightarrow 0$ at the QCP, the toy model DC conductivity matches $\sigma^{\text{DC}}(T, t_{\perp,c})$. Parameters: $U = 1.75$.}
    \label{fig:dc_cond_nrg}
\end{figure}

The DC conductivity is determined by taking the limit as $\omega \rightarrow 0$ in the optical conductivity expression (~\ref{eq:opt_cond_w}), which then reduces to the following form:

\begin{align}
\frac{\sigma^{\text{DC}}(T)}{\sigma_0} &= 2\pi \int d\epsilon \,\text{A}_0(\epsilon) \int d\omega' \left(-\frac{dn_{F}}{d\omega'}\right)\nonumber\\ &\quad\quad\quad\times \, \text{Tr}[v^2(\epsilon) \textbf{A}^2(\omega',\epsilon)]\label{eq:dc_cond}
\end{align}

DC conductivity, $\sigma^{\text{DC}}(T)$, is plotted for different inter-orbital couplings, $t_\perp$, in Fig.~\ref{fig:dc_cond_nrg}. As expected, $\sigma^{\text{DC}}(T)$ diverges for $t_\perp < t_{\perp,c}$ and vanishes for $t_\perp > t_{\perp,c}$. At the QCP, it follows a different power-law behavior. At the QCP, the characteristic $\omega/T$ scaling implies a power-law form for the self-energy, $\Sigma_{\text{f}}(\omega) \propto T^{-r}|\omega/T|^{-r}$.
This, in turn, governs the behavior of the spectral functions, resulting in the relation $\text{A}_\text{c}(\omega,\epsilon) = \text{A}_{\text{c}_M}(\omega,\epsilon) \propto T^{2r}$. This leads to the DC conductivity scaling as $\sigma_{\text{DC}} \propto T^{2r}$ at the QCP. A crude estimate of the DC conductivity can thus be obtained from Eq.~\ref{eq:dc_cond} by assuming a toy model self-energy form, $\Sigma_{\text{f}}(\omega) \propto T^{-r}|\omega/T|^{-r}$. Such a toy model calculation suggests that very low temperatures, as low as $T \sim 10^{-5}$, are required to observe this scaling clearly.
However, such calculations suffer from oscillations at very low frequencies and temperatures~\cite{vzitko2015repulsive}. Ideally, accurate results would be obtained in the limit $\Lambda \rightarrow 1$ with a larger $n_{\text{states}}$, but this remains infeasible with current computational resources.

\begin{figure}[htbp]
    \centering
    \includegraphics[width=0.85\linewidth,trim={0 0 0 1},clip]{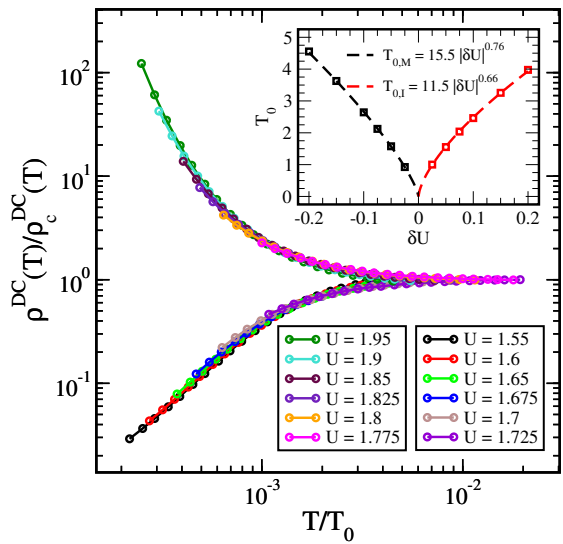}
    \caption{(Main) Scaled DC resistivity, $\rho^{DC}(T)/\rho^{DC}_c(T)$ is plotted as a function of scaled temperature($T/T_0$) for different values of $U$. (Inset) $T_0$ is plotted as a function of $\delta U=|U-U_C|$, where $U_c=1.75$. $T_{0,\text{M}},T_{0,\text{I}}$ corresponds to the scaled temperatures $T_0$ in Metal and Mott insulating side. $\text{T}_{0,\text{M}},T_{0,\text{I}}$ vanishes with exponents ($0.76, 0.66$). Parameters used: $t_\perp=t_{\perp\text{c}}=1.033$.}
    \label{fig:dc_resistivity_scaling}
\end{figure}

Resistivity is obtained from the inverse of the DC conductivity, $\rho^{\text{DC}}=1/\sigma^{\text{DC}}$.
Systems exhibiting Mott quantum criticality display a characteristic scaling behavior in the electrical resistivity~\cite{terletska2011quantum,vuvcivcevic2013finite,eisenlohr2019mott}. This scaling provides insights into the quantum critical region and allows for the extraction of the critical exponent $\nu z$, where $\nu$ is the correlation length exponent and $z$ is the dynamical exponent. A hallmark of Mott quantum criticality is the scaling collapse of the DC resistivity curves. This universal behavior is captured by the scaling form:
\begin{align}
\rho^{\text{DC}}(T,\delta U) = \rho^{\text{DC}}_{\text{c}}(T)\,\, f\left(\frac{T}{T_0(\delta U)}\right),
\end{align}
where $\rho_{c}^{\text{DC}}(T)$ is the resistivity measured precisely at the critical point and $f$ is a universal scaling function. The characteristic energy scale, $T_0$, depends on the deviation from the critical interaction, $\delta U = U - U_c$, via the power law $T_0 \sim |\delta U|^{\nu z}$, with $\nu z$ being the dynamical critical exponent. The SBHM exhibits such scaling above the finite temperature critical point, $T_{c}$, with an exponent $\nu z = 0.56 \pm 0.01$~\cite{terletska2011quantum} when calculated using IPT and CTQMC impurity solvers within DMFT. However, subsequent studies on the same model have reported slightly different values: DMFT+IPT calculations by Vu\v{c}i\v{c}evi\'{c} et al.~\cite{vuvcivcevic2013finite} yield $\nu z = 0.6$, while DMFT+NRG results by Eisenlohr et al.~\cite{eisenlohr2019mott} report $\nu z = 0.66 \pm 0.1$. This variation in the reported exponent is understood to arise from different methodological choices for the Widom line, the trajectory in the temperature-interaction plane used as a reference for the scaling analysis. In this paper, we choose a straight Widom line, similar to the approach discussed in Eisenlohr et al~\cite{eisenlohr2019mott}.

Since our model undergoes a continuous Mott transition, we expect a similar quantum critical scaling in the electrical resistivity. Here, we perform the scaling analysis at a fixed inter-orbital coupling $t_\perp$, while varying the Hubbard interaction $U$. Fig.~\ref{fig:dc_resistivity_scaling} shows the scaled resistivity $\rho^{\text{DC}}(T)/\rho^{\text{DC}}_c(T)$ as a function of the scaled temperature $T/T_0$ for different values of $U$. We observe a scaling collapse of the data down to the lowest temperatures, indicating that the scaling persists in the limit $T\rightarrow0$ on both the metallic and insulating sides. This behavior contrasts sharply with that of the SBHM~\cite{terletska2011quantum,vuvcivcevic2013finite}, where scaling was observed only above the critical temperature, $T>T_{c}$. An exception is the study by Eisenlohr et al~\cite{eisenlohr2019mott}, which reported resistivity scaling extending to temperatures below $T_c$. However, this scaling at $T<T_c$ was limited to the metastable insulating region. In contrast, our results reveal robust resistivity scaling on both sides of the transition as $T\rightarrow0$, suggesting that the quantum critical point governs the Mott critical transport observed at finite temperatures.

The scaling temperature \(T_0\) exhibits a power-law dependence, characterized by an exponent of approximately \(\nu z_{\text{met}} = 0.76\) on the metallic side and \(\nu z_{\text{ins}} = 0.66\) on the insulating side, as shown in the inset of Fig.~\ref{fig:dc_resistivity_scaling}.
The disparity in the exponents across the two sides may originate from the choice of the Widom line, as discussed by Eisenlohr et al.~\cite{eisenlohr2019mott}, who reported $\nu z\in(0.5,0.8)$ for different Widom line definitions in the SBHM. Our results for the MPAM fall within this range, indicating that the critical scaling of the MPAM is consistent with that of the SBHM.

This finding suggests that the MPAM and SBHM may belong to the same universality class. Consistent with this interpretation, experimental studies on systems exhibiting either first-order or continuous Mott transitions report critical exponents within these bounds. Therefore, the MPAM can be regarded as a prototypical model for exploring Mott quantum criticality.

\section{CONCLUSIONS AND OUTLOOK}
\label{sec:discussion}

In this paper, using NRG, we have demonstrated that a modified periodic Anderson model exhibits Mott quantum criticality, characterized by quantum critical scaling in resistivity with a critical exponent comparable to that of the SBHM. This suggests that the MPAM and SBHM may belong to the same universality class. Importantly, the critical temperature $T_c$ of the MPAM is zero. We have also shown that the MPAM displays $\omega/T$ scaling, a hallmark of local quantum criticality. Furthermore, we identified isosbestic points in the optical conductivity and showed that the isosbestic points vanish upon approaching the QCP. Analytical expressions for the MIR peak and the charge excitation peak in the optical conductivity were derived in terms of model parameters. We observed that the MIR peak reaches its minimum at the QCP.

A continuous Mott-like metal–insulator transition has been observed in the Falicov–Kimball model~\cite{haldar2016quantum,haldar2017quantum}, although its critical exponent $\nu z$ is approximately twice that found in the SBHM. Several attempts have been made to tune the finite-temperature critical point of the SBHM to $T=0$ using disorder~\cite{aguiar2005effects} or particle–hole asymmetry, but these have been unsuccessful. In contrast, the MPAM features a continuous, symmetry-unbroken transition at zero temperature from a Fermi liquid to a Mott insulating state, while showing a comparable critical exponent $\nu z$ as the SBHM. These findings open up several avenues for future exploration. A central question is whether our results define the universality class for Mott quantum criticality. Another key direction is to explore if a suitable parameter can tune the SBHM to a zero-temperature transition, thereby realizing the physics of the MPAM. Furthermore, investigating the role of disorder in this quantum critical system remains a compelling challenge for future study.

\acknowledgments
N.\ S.\ V. and S.\ K.\ K. acknowledge funding from JNCASR
and computational resources from the National Supercomputing Mission and JNCASR. S.\ K.\ K. acknowledges financial support from the DST-INSPIRE fellowship.\\

\appendix
\section{Analytic expressions for roots of the Green's function}
\label{sec:Appendix1}
In this section, we derive the analytical expressions for the roots obtained for the Green's function. The $f-$Green's function can be written as follows,
\begin{align*}
  G_{f}&(\omega,\epsilon) = \frac{1}{\omega-\Sigma_{\text{f}}(\omega)} \left(1+\frac{V^2}{2(\omega-\Sigma_{\text{f}}(\omega))}\times\right.\\&\left.\left[  \frac{1+\frac{V^2}{\sqrt{V^4+4(\omega-\Sigma)^2\,t_{\perp}^2}}}{\omega-\epsilon-\frac{V^2}{2(\omega-\Sigma_\text{f})}-\sqrt{\frac{V^4}{4(\omega-\Sigma_{\text{f}}(\omega))^2}+t_{\perp}^2}} \right.\right.\\&\left.\left.+ \frac{1-\frac{V^2}{\sqrt{V^4+4(\omega-\Sigma)^2\,t_{\perp}^2}}}{\omega-\epsilon-\frac{V^2}{2(\omega-\Sigma_\text{f})}+\sqrt{\frac{V^4}{4(\omega-\Sigma_{\text{f}}(\omega))^2}+t_{\perp}^2}}   \right]\right)
\end{align*}
The roots of the Green's function can be obtained by setting the denominator of the above expression to zero. Consider a Fermi liquid (FL) state renormalized by a quasiparticle weight $Z$. In this case, $\omega-\Sigma_{\text{f}}(\omega)=\omega/Z$. In the limit $\omega\gg\frac{V^2Z}{2\omega}$, the analytic expression for the roots can be derived as follows:
 \begin{align}
     \omega&-\epsilon-\frac{V^2Z}{2\omega}\pm t_\perp\sqrt{1+\left(\frac{ZV^2}{2\omega t_\perp}\right)^2}=0,\label{eq:A1}\\\omega&-\epsilon-\frac{V^2Z}{2\omega}\pm t_\perp=0,\label{eq:A2}\\
     \omega&=\frac{1}{2}\left( (\epsilon\pm t_{\perp})\pm\sqrt{(\epsilon\pm t_\perp)^2-2V^2Z} \right).
 \end{align}
There are 4 possibilities, but one can get the following 3 unique roots,
\begin{align*}
    R_u &= \frac{\epsilon + t_\perp + \sqrt{(\epsilon + t_\perp)^2 + 2ZV^2}}{2},\\
    R_m &= \frac{\epsilon - \text{sgn}(\epsilon)\, t_\perp + \text{sgn}(\epsilon)\sqrt{(\epsilon - \text{sgn}(\epsilon)t_\perp)^2 + 2ZV^2}}{2},\\
    R_l &= \frac{\epsilon - t_\perp - \sqrt{(\epsilon - t_\perp)^2 + 2ZV^2}}{2}.
\end{align*}
One can get the minimum gap between these roots by taking the derivative of the gap with respect to $\epsilon$. 

\section{Derivation of c-band spectral function}
\label{sec:Appendix2}
In this section, we derive the expression for the c-band spectral function, $\text{A}_{\text{c}}(\omega)$, in the low-frequency limit ($\omega\rightarrow0$). Consider the Green's function for the middle conduction band, as given in Eq. \ref{eq:Gcc}. Using a partial fraction decomposition, this equation can be rewritten as:
\begin{align}
    G_{\text{cc}}&(\omega,\epsilon) = \frac{1}{2} \Bigg(  \frac{1+\frac{V^2}{\sqrt{V^4+4(\omega-\Sigma)^2\,t_{\perp}^2}}}{\omega-\frac{V^2}{2(\omega-\Sigma_\text{f})}-\sqrt{\frac{V^4}{4(\omega-\Sigma_{\text{f}}(\omega))^2}+t_{\perp}^2}-\epsilon} \nonumber\\&+ \frac{1-\frac{V^2}{\sqrt{V^4+4(\omega-\Sigma)^2\,t_{\perp}^2}}}{\omega-\frac{V^2}{2(\omega-\Sigma_\text{f})}+\sqrt{\frac{V^4}{4(\omega-\Sigma_{\text{f}}(\omega))^2}+t_{\perp}^2}-\epsilon}   \Bigg)
\end{align}
In the Fermi-liquid regime, we employ a similar argument to that presented in Appendix~\ref{sec:Appendix1} to make the low-frequency approximation $(\omega-\Sigma_\text{f})\approx\omega/Z$. Substituting this into the expression for $G_{\text{cc}}(\omega,\epsilon)$ yields:
\begin{align}
    G_{\text{cc}}(\omega,\epsilon) = \frac{1}{2}\Bigg( \frac{1+ \left(1+\frac{4\omega^2t_{\perp}^2}{Z^2V^4}\right)^{-\frac{1}{2}}}{\omega-\frac{ZV^2}{2\omega}-\sqrt{\frac{Z^2V^4}{4\omega^2}+t_{\perp}^2}-\epsilon} \nonumber\\
    \qquad+ \frac{1-\left(1+\frac{4\omega^2t_{\perp}^2}{Z^2V^4}\right)^{-\frac{1}{2}}}{\omega-\frac{ZV^2}{2\omega}+\sqrt{\frac{Z^2V^4}{4\omega^2}+t_{\perp}^2}-\epsilon}\Bigg)\label{eq:Gcc_FL}
\end{align}
The c-band spectral function is calculated from the expression $\displaystyle\text{A}_{\text{c}}(\omega)=-\frac{1}{\pi}\text{Im}\Bigg[\int d\epsilon\, \text{A}_0(\epsilon) \,G_{\text{cc}}(\omega,\epsilon)\Bigg]$. In the $\omega\rightarrow0$ limit, simplifying Eq.~\ref{eq:Gcc_FL} by using the approximation $(1+x)^{-1/2}\approx1-x/2,\, \forall \,x\ll1$. Substituting this result into the integral for $\displaystyle\text{A}_{\text{c}}(\omega)$ yields the final form of the c-band spectral function presented in Subsection \ref{subsecA}.

\bibliography{reference}
\end{document}